# Non-stationary longitudinal Josephson effect in electron-hole bilayers

O.M. Konstantynov, S.I. Shevchenko

*B. Verkin Institute for Low Temperature Physics and Engineering of NAS of Ukraine,*

*47 Nauky Ave., Kharkiv, 61103, Ukraine*

E-mail: akonstantinov@ilt.kharkov.ua

The non-stationary longitudinal Josephson effect in bilayer systems with pairing of spatially separated electrons and holes is investigated. The tunneling current between two (right and left) weakly coupled electron-hole condensates is analyzed in the presence of an external voltage applied along the layers. Two limiting regimes are considered: the high-density regime, where pairing occurs in momentum space similarly to conventional superconductors, and the low-density regime corresponding to a Bose-Einstein condensate of spatially indirect excitons. Using the tunneling Hamiltonian approach, expressions for the quasiparticle, superconducting and interference contributions to the tunneling current are derived. It is shown that in the high-density limit the quasiparticle and interference currents vanish at zero temperature below a threshold voltage determined by the energy gap, so that only the supercurrent remains. In contrast, in the low-density regime the gapless spectrum of collective excitations leads to finite dissipative contributions even at zero temperature, resulting in damping of Josephson oscillations. For spatially separated tunneling barriers with distance exceeding the coherence length, phase jumps occur at each barrier, resulting in voltage oscillations across the electron and hole barriers and a non-sinusoidal current-phase relation governed by the critical currents of the two barriers.

## 1. Introduction

The counterflow superconductivity predicted about fifty years ago in systems with pairing of spatially separated electrons and holes continues to be a subject of active theoretical [1-26] and experimental studies [27-33]. In particular, in [21] the features of the longitudinal stationary Josephson effect between the left and right condensates in such systems were considered and it was shown that the results depend significantly on the carrier density $n_e = n_h = n$. In the low-density limit ($na_B^2 \ll 1$ , where $a_B$ is the average pair size), the system represents a gas of spatially indirect excitons whose behavior is similar to that of a Bose gas. In the high-density regime ( $na_B^2 \gg 1$), the specific forms of the electron and hole dispersion laws begin to play a key role. It was shown long ago [34] that for congruent electron and hole dispersion laws pairing of carriers occurs in momentum space – analogously to what takes place in conventional superconductors – independently of the interaction strength and the screening radius. Although in the work [34] systems without spatial separation of carriers were considered, the results obtained there remain valid for bilayer electron-hole systems [21]. If the dispersion laws are non-congruent, the system undergoes a Mott transition as the density increases [35] and the exciton gas transforms into an electron-hole liquid. This circumstance significantly restricts the region of existence of the exciton condensate in the high-density regime and requires a more careful choice of model systems. Nevertheless, there are physically realizable bilayer systems in which the congruence of the carrier Fermi surfaces is fulfilled with high accuracy, and the threshold carrier density approaches $10^{13}\,cm^{-2}$ [36]. This makes the analysis of the high-density regime justified and physically significant, for example, for studying the BEC-BCS crossover.

In the present work, we consider the non-stationary longitudinal Josephson effect in bilayer systems with pairing of spatially separated electrons and holes in the presence of an external voltage applied between the left and right sides of the system. The analysis is performed both in the low-density limit corresponding to an excitonic Bose condensate and in the high-density limit where electron-hole pairing is realized in momentum space.

## 2. High density limit

First, let us consider the problem of longitudinal tunneling in a bilayer system with spatially separated electrons and holes in the high-density limit so that $na_B^2 >> 1$.

The tunnel junction separating the system into the left and right regions (see Fig. 1) is described by a Hamiltonian of the form

$$H = H_L + H_R + H_T, \tag{1}$$

where $H_L$ and $H_R$ are the full Hamiltonians of the left and right half-spaces corresponding to isolated bilayer systems, while the term $H_T$ describes the transfer of electrons and holes through the barrier:

$$H_T = \sum_{\mathbf{p},\mathbf{q},\sigma} \left( T^e_{\mathbf{p},\mathbf{q}} a^{\dagger}_{1\mathbf{p},\sigma} a_{2\mathbf{q},\sigma} + T^h_{\mathbf{p},\mathbf{q}} b^{\dagger}_{1\mathbf{p},\sigma} b_{2\mathbf{q},\sigma} + h.c. \right), \tag{2}$$

where $a^{\dagger}_{1\mathbf{p},\sigma}$ and $b^{\dagger}_{1\mathbf{p},\sigma}$ are the creation operators for an electron and a hole, respectively, with momentum $\mathbf{p}$ and spin projection $\sigma$ to the left of the barrier, while $a^{\dagger}_{2\mathbf{q},\sigma}$ and $b^{\dagger}_{2\mathbf{q},\sigma}$ are the corresponding operators for electrons and holes on the right side of the barrier. The quantities $T^e_{\mathbf{p},\mathbf{q}}$ and $T^h_{\mathbf{p},\mathbf{q}}$ are the matrix elements determining the probability of electron and hole tunneling, respectively.

*Fig. 1. Schematic representation of a bilayer system in which the left and right regions with electron-hole pairing are weakly coupled.*

If a total voltage $V(t) = V_e(t) + V_h(t)$ is applied to the contact (a voltage $V_e(t)$ in the electron layer and a voltage $V_h(t)$ in the hole layer – see Fig. 1), the Fermi levels of electrons and holes are shifted by $eV_{e(h)}(t)$ ($e$ is the elementary electric charge), which is equivalent to the presence of an additional energy for the electrons and holes in the left region. As a result, the Hamiltonian $H_L$ can be written in the form

$$H_L(V) = H_L(0) - eV_e N^e_L + e(-V_h) N^h_L, \tag{3}$$

where $N^e_L$ and $N^h_L$ are the particle-number operators for electrons and holes in the left half-space, defined by the following expressions

$$N^e_L = \sum_{\mathbf{p},\sigma} a^{\dagger}_{1\mathbf{p},\sigma} a_{1\mathbf{p},\sigma}, \; N^h_L = \sum_{\mathbf{p},\sigma} b^{\dagger}_{1\mathbf{p},\sigma} b_{1\mathbf{p},\sigma}. \tag{4}$$

The second and third terms in (3) may formally be eliminated by a time dependent gauge transformation:

$$a_{1\mathbf{p},\sigma}(t) \to \exp\left( \frac{i\varphi_e(t)}{2} \right) a_{1\mathbf{p},\sigma}(t), \; b_{1\mathbf{p},\sigma}(t) \to \exp\left( \frac{i\varphi_h(t)}{2} \right) b_{1\mathbf{p},\sigma}(t), \tag{5}$$

where

$$\frac{d\varphi_{e,h}(t)}{dt} = \frac{2e}{\hbar} V_{e,h}(t). \tag{6}$$

Note that in the presence of the voltage $V(t)$ the form of the Hamiltonian $H_R$ remains unchanged and, consequently, the operators corresponding to the right region $a_{2\mathbf{q},\sigma}$ and $b_{2\mathbf{q},\sigma}$ also remain unchanged.

The tunneling current, for example, in the electron layer $I_e(t)$ is determined by the rate of change of the number of electrons in one of the half-spaces (see, e.g., [37])

$$I_e = e\left\langle \frac{dN_L^e}{dt} \right\rangle = \frac{2e}{\hbar} \mathrm{Im} \sum_{\mathbf{p},\mathbf{q},\sigma} T_{\mathbf{p},\mathbf{q}}^e \left\langle a_{1\mathbf{p},\sigma}^\dagger a_{2\mathbf{q},\sigma} \right\rangle \exp\left( -\frac{i\varphi_e(t)}{2} \right), \tag{7}$$

where averaging is performed over the state of the full Hamiltonian. To calculate the average appearing in (7) we pass from the Heisenberg representation to the interaction representation and use perturbation theory, treating the term $H_T$ in the Hamiltonian as a perturbation. As a result, to first order in $H_T$ we obtain

$$I_e = -\frac{2e}{\hbar^2} \mathrm{Re} \sum_{\mathbf{p},\mathbf{q},\sigma} T_{\mathbf{p},\mathbf{q}}^e \exp\left( -\frac{i\varphi_e(t)}{2} \right) \int_{-\infty}^{t} d\tau e^{\eta\tau} \left\langle \left[ a_{1\mathbf{p},\sigma}^\dagger a_{2\mathbf{q},\sigma}, H_T(\tau) \right] \right\rangle_0, \tag{8}$$

where $\eta \to 0^+$, and the symbol $\langle ... \rangle_0$ denotes averaging with respect to the state of the unperturbed Hamiltonian $H_0 = H_L + H_R$. Substituting the tunneling Hamiltonian (2) into this expression and taking into account that averages of the type $\left\langle a_{1\mathbf{p},\sigma}^\dagger a_{1\mathbf{p}',\sigma'}^\dagger \right\rangle_0$ and $\left\langle a_{1\mathbf{p},\sigma}^\dagger b_{1\mathbf{p}',\sigma'} \right\rangle_0$ vanish, we obtain

$$I_e = \mathrm{Im} \int_{-\infty}^{+\infty} d\tau e^{\eta\tau} \left\{ e^{-\frac{i}{2}[\varphi_e(t) - \varphi_e(\tau)]} S(t-\tau) + e^{-\frac{i}{2}[\varphi_e(t) + \varphi_h(\tau)]} R(t-\tau) \right\}. \tag{9}$$

Here the function $S(t-\tau)$ is expressed in terms of the single-particle Green's functions $G$, while $R(t-\tau)$ is expressed through the anomalous Green's functions $F$ and $\tilde{F}$ as follows:

$$S(t-\tau) = -4i\frac{e}{\hbar^2}\theta(t-\tau)\sum_{\mathbf{p},\mathbf{q}} |T_{\mathbf{pq}}^e|^2 \left\{ G^<(\mathbf{p},\tau-t)G^>(\mathbf{q},t-\tau) - G^<(\mathbf{q},t-\tau)G^>(\mathbf{p},\tau-t) \right\}, \tag{10}$$

$$R(t-\tau) = -4i\frac{e}{\hbar^2}\theta(t-\tau)\sum_{\mathbf{p},\mathbf{q}} T_{\mathbf{pq}}^e T_{\mathbf{pq}}^{h\,*} \left\{ \tilde{F}^>(\mathbf{p},t-\tau)F^<(\mathbf{q},\tau-t) - \tilde{F}^<(\mathbf{p},t-\tau)F^>(\mathbf{q},\tau-t) \right\}. \tag{11}$$

Here the summation over spin indices has already been performed, and the relation $T_{-\mathbf{p},-\mathbf{q}}^{e(h)} = T_{\mathbf{p},\mathbf{q}}^{e(h)*}$ has been used. Note that the Green's functions entering (10) and (11) are not time-ordered, as indicated by the superscript $\gtrless$, namely

$$\begin{aligned}
G^>(\mathbf{p},\tau-t) &= -i\left\langle a_{1\mathbf{p}\uparrow}(\tau)a_{1\mathbf{p}\uparrow}^\dagger(t) \right\rangle_0, \; G^<(\mathbf{p},\tau-t) = i\left\langle a_{1\mathbf{p}\uparrow}^\dagger(t)a_{1\mathbf{p}\uparrow}(\tau) \right\rangle_0, \\
G^>(\mathbf{q},t-\tau) &= -i\left\langle a_{2\mathbf{q}\uparrow}(t)a_{2\mathbf{q}\uparrow}^\dagger(\tau) \right\rangle_0, \; G^<(\mathbf{q},t-\tau) = i\left\langle a_{2\mathbf{q}\uparrow}^\dagger(\tau)a_{2\mathbf{q}\uparrow}(t) \right\rangle_0, \\
\tilde{F}^>(\mathbf{p},t-\tau) &= \left\langle a_{1\mathbf{p}\uparrow}^\dagger(t)b_{1-\mathbf{p}\downarrow}^\dagger(\tau) \right\rangle_0, \; \tilde{F}^<(\mathbf{p},t-\tau) = -\left\langle b_{1-\mathbf{p}\downarrow}^\dagger(\tau)a_{1\mathbf{p}\uparrow}^\dagger(t) \right\rangle_0, \\
F^>(\mathbf{q},\tau-t) &= \left\langle b_{2-\mathbf{q}\downarrow}(\tau)a_{2\mathbf{q}\uparrow}(t) \right\rangle_0, \; F^<(\mathbf{q},\tau-t) = -\left\langle a_{2\mathbf{q}\uparrow}(t)b_{2-\mathbf{q}\downarrow}(\tau) \right\rangle_0.
\end{aligned} \tag{12}$$

The Fourier components of the Green's functions can be expressed in terms of the corresponding spectral functions $A(\mathbf{k},\omega)$ and $B(\mathbf{k},\omega)$

$$G^{\gtrless}(\mathbf{p},\omega) = \mp iA(\mathbf{p},\omega)f^\pm(\omega), \tag{13}$$

$$\tilde{F}^{\gtrless}(\mathbf{p},\omega) = \pm e^{-i\varphi_L} B(\mathbf{p},\omega)f^\pm(\omega), \tag{14}$$

$$F^{\gtrless}(\mathbf{p},\omega) = \pm e^{i\varphi_R} B(\mathbf{q},\omega)f^\pm(\omega), \tag{15}$$

where $f^\pm(\omega) = (\exp(\mp\omega\hbar/T)+1)^{-1}$, and $T$ is the temperature expressed in energy units. In expressions (14) and (15) , the phase factors in the spectral functions $B$ have been explicitly separated. As a result, expressions (10) and (11) can be rewritten in the form

$$S(t) = 4i\frac{e}{\hbar^2}\theta(t)\sum_{\mathbf{p},\mathbf{q}} \left|T_{\mathbf{pq}}^e\right|^2 \int_{-\infty}^{+\infty}\int_{-\infty}^{+\infty} \frac{d\omega d\omega'}{(2\pi)^2} e^{i(\omega-\omega')t} A(\mathbf{p},\omega)A(\mathbf{q},\omega')\left[ f^-(\omega') - f^-(\omega) \right], \tag{16}$$

$$\tilde{R}(t) = 4i\frac{e}{\hbar^2}\theta(t)\sum_{\mathbf{p},\mathbf{q}} T^e_{\mathbf{pq}}T^{h*}_{\mathbf{pq}}\int_{-\infty}^{+\infty}\int_{-\infty}^{+\infty}\frac{d\omega d\omega'}{(2\pi)^2}e^{-i(\omega-\omega')t}B(\mathbf{p},\omega)B(\mathbf{q},\omega')\left[f^-(\omega')-f^-(\omega)\right]. \quad (17)$$

Here a new function $\tilde{R}(t) = \exp(i(\varphi_L - \varphi_R))R(t)$ is introduced. If we introduce the spectral decomposition

$$e^{-\frac{i\varphi_{e(h)}(t)}{2}} \equiv \int_{-\infty}^{+\infty}\frac{d\omega}{2\pi}W_{e(h)}(\omega)e^{-i\omega t}, \quad (18)$$

then the current (9) can be rewritten as

$$I_e = \mathrm{Im}\left(\int_{-\infty}^{+\infty}\int_{-\infty}^{+\infty}\frac{d\omega d\omega'}{(2\pi)^2}W_e(\omega)\left\{e^{-i(\omega-\omega')t}W_e^*(\omega')S(i\eta-\omega')+e^{-i(\omega+\omega')t}e^{-i\varphi_0}W_h(\omega')\tilde{R}(i\eta+\omega')\right\}\right). \quad (19)$$

Here $\varphi_0 = \varphi_L - \varphi_R$ and functions $S(\omega)$ and $\tilde{R}(\omega)$ are the Fourier transforms of the functions $S(t)$ and $\tilde{R}(t)$, respectively. In the case of a constant external voltage $V(t) = V_0 = const$ and assuming that $V_{e(h)} \neq V_{e(h)}(t)$, we have

$$W_{e(h)}(\omega) = 2\pi\delta\left(\omega - \frac{eV_{e,h}}{\hbar}\right). \quad (20)$$

From (19) and (20) we obtain the following expression for the tunneling current:

$$I_e = I_q\left(\frac{eV_e}{\hbar}\right) + I_{J1}\left(\frac{eV_h}{\hbar}\right)\sin\left(\frac{eV_0}{\hbar}t+\varphi_0\right) + I_{J2}\left(\frac{eV_h}{\hbar}\right)\cos\left(\frac{eV_0}{\hbar}t+\varphi_0\right). \quad (21)$$

The quasiparticle current $I_q$ entering (21), as well as the amplitudes of the supercurrent $I_{J1}$ and the current $I_{J2}$, arising from the interference of the supercurrent and quasiparticle current (hereafter referred to as the interference current), can be expressed in terms of the Fourier transforms of the functions $S(t)$ and $\tilde{R}(t)$:

$$I_q(\omega_0) = \mathrm{Im}\,S(i\eta-\omega_0),\; I_{J1}(\omega_0) = -\mathrm{Re}\,\tilde{R}(i\eta+\omega_0),\; I_{J2}(\omega_0) = \mathrm{Im}\,\tilde{R}(i\eta+\omega_0), \quad (22)$$

where $\omega_0$ denotes the argument of the current amplitudes appearing in expression (21). Using expressions (16) and (17), we obtain

$$\begin{aligned} S(i\eta-\omega_0) = \frac{4e}{\hbar^2}\sum_{\mathbf{p},\mathbf{q}}\left|T^e_{\mathbf{pq}}\right|^2 P\int_{-\infty}^{+\infty}\int_{-\infty}^{+\infty}\frac{d\omega d\omega'}{(2\pi)^2}A(\mathbf{p},\omega)A(\mathbf{q},\omega')\left[f^-(\omega)-f^-(\omega')\right]\times \\ \times\left\{\frac{1}{\omega-\omega'-\omega_0} - i\pi\delta(\omega-\omega'-\omega_0)\right\}, \end{aligned} \quad (23)$$

$$\begin{aligned} \tilde{R}(i\eta+\omega_0) = \frac{4e}{\hbar^2}\sum_{\mathbf{p},\mathbf{q}}T^e_{\mathbf{pq}}T^{h*}_{\mathbf{pq}}P\int_{-\infty}^{+\infty}\int_{-\infty}^{+\infty}\frac{d\omega d\omega'}{(2\pi)^2}B(\mathbf{p},\omega)B(\mathbf{q},\omega')\left[f^-(\omega')-f^-(\omega)\right]\times \\ \times\left\{\frac{1}{\omega-\omega'-\omega_0} + i\pi\delta(\omega-\omega'-\omega_0)\right\}. \end{aligned} \quad (24)$$

where the symbol $P$ indicates that the remaining integral should be evaluated in the sense of the principal value. Substituting (23), (24) into (22), we obtain the following expressions for the current amplitudes

$$I_q(\omega_0) = \frac{e}{\hbar^2\pi}\sum_{\mathbf{p},\mathbf{q}}\left|T^e_{\mathbf{pq}}\right|^2\int_{-\infty}^{+\infty}d\omega A(\mathbf{p},\omega)A(\mathbf{q},\omega-\omega_0)\left[f^-(\omega-\omega_0)-f^-(\omega)\right], \quad (25)$$

$$I_{J1}(\omega_0) = -\frac{4e}{\hbar^2}\sum_{\mathbf{p},\mathbf{q}}T^e_{\mathbf{pq}}T^{h*}_{\mathbf{pq}}P\int_{-\infty}^{+\infty}\int_{-\infty}^{+\infty}\frac{d\omega d\omega'}{(2\pi)^2}\frac{B(\mathbf{p},\omega)B(\mathbf{q},\omega')}{\omega-\omega'-\omega_0}\left[f^-(\omega')-f^-(\omega)\right], \quad (26)$$

$$I_{J2}(\omega_0) = \frac{e}{\hbar^2\pi}\sum_{\mathbf{p},\mathbf{q}}T^e_{\mathbf{pq}}T^{h*}_{\mathbf{pq}}\int_{-\infty}^{+\infty}d\omega B(\mathbf{p},\omega)B(\mathbf{q},\omega-\omega_0)\left[f^-(\omega-\omega_0)-f^-(\omega)\right]. \quad (27)$$

Calculating the spectral functions $A(\mathbf{k},\omega)$ and $B(\mathbf{k},\omega)$ similarly to the case of a conventional superconductor (see, e.g., [38]), we find

$$A(k,\omega)=\pi\left[\left(1-\frac{\xi_k}{\varepsilon_k}\right)\delta\left(\omega-\frac{\varepsilon_k-\eta_k}{\hbar}\right)-\left(1+\frac{\xi_k}{\varepsilon_k}\right)\delta\left(\omega-\frac{\varepsilon_k+\eta_k}{\hbar}\right)\right], \tag{28}$$

$$B(k,\omega)=\frac{\pi\Delta}{\varepsilon_k}\left[\delta\left(\omega-\frac{\varepsilon_k-\eta_k}{\hbar}\right)-\delta\left(\omega+\frac{\varepsilon_k+\eta_k}{\hbar}\right)\right], \tag{29}$$

where $\varepsilon_k=\sqrt{\xi_k^2+\Delta^2}$, $\xi_k=\left(\xi_k^e+\xi_k^h\right)/2$, $\eta_k=\left(\xi_k^e-\xi_k^h\right)/2$, $\xi_k^e$ and $\xi_k^h$ – are the energies of the electron and the hole, measured from their respective chemical potentials, and $\Delta$ – is the energy gap. Note that for equal carrier masses the form of the spectral functions $A(k,\omega)$ and $B(k,\omega)$ coincides with that in ordinary superconductors. Substituting (28) and (29) into (25)-(27), and assuming for definiteness that $\omega_0>0$, we finally obtain the following expressions for the quasiparticle current

$$\begin{aligned}I_q(\omega_0)=\frac{e\pi}{\hbar}\sum_{\mathbf{p},\mathbf{q}}\left|T_{\mathbf{pq}}^e\right|^2&\left\{\left(1+\frac{\xi_p}{\varepsilon_p}\right)\left(1+\frac{\xi_q}{\varepsilon_q}\right)\left(f_q^e-f_p^e\right)\delta\left(\hbar\omega_0+\varepsilon_q^e-\varepsilon_p^e\right)+\right.\\&+\left(1-\frac{\xi_p}{\varepsilon_p}\right)\left(1-\frac{\xi_q}{\varepsilon_q}\right)\left(f_p^h-f_q^h\right)\delta\left(\hbar\omega_0+\varepsilon_p^h-\varepsilon_q^h\right)+\\&\left.+\left(1+\frac{\xi_p}{\varepsilon_p}\right)\left(1-\frac{\xi_q}{\varepsilon_q}\right)\left(1-f_p^e-f_q^h\right)\delta\left(\hbar\omega_0-\varepsilon_p^e-\varepsilon_q^h\right)\right\},\end{aligned} \tag{30}$$

for the amplitude of the supercurrent

$$\begin{aligned}I_{J1}(\omega_0)=-\frac{e\Delta^2}{\hbar}\sum_{\mathbf{p},\mathbf{q}}\frac{T_{\mathbf{pq}}^eT_{\mathbf{pq}}^{h*}}{\varepsilon_p\varepsilon_q}&\left\{\frac{f_q^e-f_p^e}{\varepsilon_p^e-\varepsilon_q^e+\hbar\omega_0}+\frac{f_q^h-f_p^h}{\varepsilon_p^h-\varepsilon_q^h-\hbar\omega_0}+\right.\\&\left.+\frac{f_p^e+f_q^h-1}{\varepsilon_p^e+\varepsilon_q^h+\hbar\omega_0}+\frac{f_p^h+f_q^e-1}{\varepsilon_p^h+\varepsilon_q^e-\hbar\omega_0}\right\},\end{aligned} \tag{31}$$

and for the amplitude of the interference current

$$\begin{aligned}I_{J2}(\omega_0)=-\frac{e\Delta^2\pi}{\hbar}\sum_{\mathbf{p},\mathbf{q}}\frac{T_{\mathbf{pq}}^eT_{\mathbf{pq}}^{h*}}{\varepsilon_p\varepsilon_q}&\left\{\left(f_q^h-f_p^h\right)\delta\left(\hbar\omega_0+\varepsilon_q^h-\varepsilon_p^h\right)+\right.\\&\left.+\left(f_p^e-f_q^e\right)\delta\left(\hbar\omega_0+\varepsilon_p^e-\varepsilon_q^e\right)+\left(f_p^h+f_q^e-1\right)\delta\left(\hbar\omega_0-\varepsilon_q^e-\varepsilon_p^h\right)\right\}.\end{aligned} \tag{32}$$

Here $\varepsilon_p^e=\varepsilon_p+\eta_p$, $\varepsilon_p^h=\varepsilon_p-\eta_p$, $f_p^e\equiv f^-(\varepsilon_p^e/\hbar)$, $f_p^h\equiv f^-(\varepsilon_p^h/\hbar)$. From (30) and (32) it follows that at zero temperature the expressions for the quasiparticle current $I_q$ and the amplitude of the interference current $I_{J2}$ take the form

$$I_q\left(\frac{eV_e}{\hbar}\right)=\frac{e\pi}{\hbar}\sum_{\mathbf{p},\mathbf{q}}\left|T_{\mathbf{pq}}^e\right|^2\left(1+\frac{\xi_p}{\varepsilon_p}\right)\left(1-\frac{\xi_q}{\varepsilon_q}\right)\delta\left(\varepsilon_p^e+\varepsilon_q^h-eV_e\right), \tag{33}$$

$$I_{J2}\left(\frac{eV_h}{\hbar}\right)=\frac{e\Delta^2\pi}{\hbar}\sum_{\mathbf{p},\mathbf{q}}\frac{T_{\mathbf{pq}}^eT_{\mathbf{pq}}^{h*}}{\varepsilon_p\varepsilon_q}\delta\left(\varepsilon_q^e+\varepsilon_p^h-eV_h\right). \tag{34}$$

These expressions show that at $T=0$ both the quasiparticle and the interference currents vanish for voltages smaller than the threshold value $V_t=2\Delta/e$ and only the supercurrent flows in the system. It should also be noted that in the case $V_e\neq V_h$ a situation may arise in which, for example, the interference current is present in the system while the quasiparticle current is equal to zero, which would correspond to the condition $V_e<V_t<V_h$.

It should be emphasized, however, that due to the asymmetry of the electron and hole dispersion relations with respect to their effective masses, as well as due to the asymmetry of the barriers in the electron and hole layers, a situation can occur in which the current in the electron layer $I_e$ at a given moment of time differs from the current in the hole layer

$$I_h = I_q^{(h)}\left(\frac{eV_h}{\hbar}\right) + I_{J1}^{(h)}\left(\frac{eV_e}{\hbar}\right)\sin\left(\frac{eV_0}{\hbar}t + \varphi_0\right) + I_{J2}^{(h)}\left(\frac{eV_e}{\hbar}\right)\cos\left(\frac{eV_0}{\hbar}t + \varphi_0\right), \quad (35)$$

where the amplitudes $I_q^{(h)}$, $I_{J1}^{(h)}$ and $I_{J2}^{(h)}$ coincide with the corresponding amplitudes $I_q$, $I_{J1}$ and $I_{J2}$ up to the interchange of the indices $e \leftrightarrow h$. Since the tunnel junctions simultaneously act as capacitors possessing a finite capacitance $C_{e(h)}$, this leads to charge accumulation on them and to the appearance of displacement currents $I_{De(h)} = (C_{e(h)}\hbar / 2e) d^2\varphi_{e(h)} / dt^2$. As a result, from the current continuity condition we obtain for the relative phase $\varphi_{e(h)}$ a dynamical equation $I_{De} + I_e = I_{Dh} + I_h$, supplemented by the condition $\varphi_e + \varphi_h = eV_0 t / \hbar + \varphi_0$ (similar to the resistively and capacitively shunted Josephson junction (RCSJ) model for two series Josephson junctions in a conventional superconductor under a fixed voltage $V_0$ [39]). In this case, the solutions (21) and (35) for the currents $I_e$ and $I_h$ in the electron and hole layers, respectively, can be regarded as quasistatic, corresponding to the adiabatic approximation [40].

### 3. Low density limit

The superfluid state of a dilute electron-hole gas ($na_B^2 << 1$) can be described using a generalized coherent state proposed by Keldysh [41]. The corresponding Keldysh wave function of the ground state of the system has the form (see also Appendix)

$$\left|\Phi_K\right\rangle = \exp\left\{ \sum_{\sigma,\sigma'=\uparrow,\downarrow} \int d\mathbf{r}_1 d\mathbf{r}_2 \Phi_{\sigma\sigma'}(\mathbf{r}_1,\mathbf{r}_2) e^{-i\mu_{\sigma\sigma'}t/\hbar} \psi_\sigma^{(e)\dagger}(\mathbf{r}_1) \psi_{\sigma'}^{(h)\dagger}(\mathbf{r}_2) - h.c. \right\} \left|0\right\rangle, \quad (36)$$

where $\psi_\sigma^{(e)\dagger}$ and $\psi_\sigma^{(h)\dagger}$ are the field creation operators for the electron and the hole, respectively, $\sigma$ is the spin index, and $\mu_{\sigma\sigma'}$ is the chemical potential of the electron-hole gas described by the function (order parameter) $\Phi_{\sigma\sigma'}(\mathbf{r}_1,\mathbf{r}_2)$. In what follows, we assume that the carrier pairing is singlet, and also that the system has a phase separation such that in the matrix $\Phi_{\sigma\sigma'}(\mathbf{r}_1,\mathbf{r}_2)$ only, for example, the component $\Phi_{\uparrow\downarrow}(\mathbf{r}_1,\mathbf{r}_2) \equiv \Phi(\mathbf{r}_1,\mathbf{r}_2)$ is nonzero. The equation for the function $\Phi(\mathbf{r}_1,\mathbf{r}_2)$ is obtained from the condition that the free-energy functional $F \equiv E - \mu N$ is minimized, where $E$ and $N$ are the energy of the system and the number of electron-hole pairs, respectively, and $\mu \equiv \mu_{\uparrow\downarrow}$. The solution of this equation is sought in the form $\Phi(\mathbf{r}_1,\mathbf{r}_2) = \Psi\left(\mathbf{R}_{12}\right)\phi\left(\mathbf{r}_1,\mathbf{r}_2\right)$, where $\mathbf{R}_{12}$ is the center-of-mass coordinate, $\Psi\left(\mathbf{R}_{12}\right)$ is the wave function of the pair moving as whole, and $\phi\left(\mathbf{r}_1,\mathbf{r}_2\right)$ is a function describing the bound state of the electron and hole. As such a function, the wave function of the ground state of the Schrödinger equation for a single electron-hole pair is taken, i.e. $\phi\left(\mathbf{r}_1,\mathbf{r}_2\right) = \phi_0\left(r_{12}\right)$ ($r_{12}$ is the relative coordinate of the electron-hole pair). The equation for the function $\Psi\left(\mathbf{R}_{12}\right)$ has the form [12]

$$\begin{aligned} &\left[ -\frac{\hbar^2}{2m}\frac{\partial^2}{\partial \mathbf{R}_{12}^2} - \tilde{\mu} + U_e\left(\mathbf{R}_{12}\right) + U_h\left(\mathbf{R}_{12}\right) \right] \Psi(\mathbf{R}_{12}) + \\ &+ \int V_d\, \phi_0^2\left(r_{12}\right) \phi_0^2\left(r_{34}\right) \Psi\left(\mathbf{R}_{12}\right) \left|\Psi\left(\mathbf{R}_{34}\right)\right|^2 d\mathbf{r}_3 d\mathbf{r}_4 d\mathbf{r}_{12} - \\ &- \int V_{ex}\, \phi_0\left(r_{12}\right) \phi_0\left(r_{23}\right) \phi_0\left(r_{34}\right) \phi_0\left(r_{14}\right) \Psi\left(\mathbf{R}_{32}\right) \Psi^*\left(\mathbf{R}_{34}\right) \Psi\left(\mathbf{R}_{14}\right) d\mathbf{r}_3 d\mathbf{r}_4 d\mathbf{r}_{12} = 0. \end{aligned} \quad (37)$$

Here $m = m_e + m_h$ is the sum of the effective masses of the electron and the hole, $\tilde{\mu}$ is the chemical potential measured from the ground-state energy of an isolated pair, $U_\alpha(\mathbf{r})$ is the external potential in the electron ($\alpha = e$) and hole ($\alpha = h$) layers, $V_d$ and $V_{ex}$ are the potentials of the direct and exchange interactions of the carriers. In deriving this equation, it has been assumed that the characteristic length scales over which the potentials $U_\alpha(\mathbf{r})$ vary are much larger than the length scales over which the function $\phi_0(r)$ changes. Using also the fact that the characteristic length scales over which the function $\Psi$ varies are much larger than both the atomic length scales and the interlayer distance $d$, Eq. (37) can be reduced to the Gross-Pitaevskii equation

$$\left[-\frac{\hbar^2}{2M}\frac{\partial^2}{\partial \mathbf{R}^2}-\tilde{\mu}+U_e(\mathbf{R})+U_h(\mathbf{R})\right]\Psi(\mathbf{R})+\gamma\,\Psi(\mathbf{R})\left|\Psi(\mathbf{R})\right|^2=0. \tag{38}$$

Here $\gamma=\gamma_d+\gamma_{ex}$ is the interaction constant, which contains contributions from the direct interaction

$$\gamma_d=\frac{4\pi e^2 d}{\varepsilon}, \tag{39}$$

and the exchange interaction

$$\gamma_{ex}=-\frac{4\pi e^2}{\varepsilon}\int\frac{d\mathbf{p}d\mathbf{q}}{(2\pi)^4}\frac{\left|\phi_\mathbf{q}\right|^2}{p}\left[\left|\phi_{\mathbf{q}+\mathbf{p}}\right|^2-\frac{e^{-pd}}{2}\left(\phi^*_{\mathbf{q}+\mathbf{p}}\phi_\mathbf{q}+\phi^*_\mathbf{q}\phi_{\mathbf{q}+\mathbf{p}}\right)\right], \tag{40}$$

where $\phi_\mathbf{q}$ is the Fourier component of the wave function $\phi_0(r)$.

In the presence of a tunnel junction separating the system into left and right regions with electron-hole pairing (see Fig. 1), a longitudinal Josephson effect will occur in the system. Its description is analogous to that in the case of a high carrier density and can be carried out using the tunneling Hamiltonian (1), in which the fermionic creation and annihilation operators are replaced by the creation $c^\dagger_\mathbf{p}$ and annihilation $c_\mathbf{p}$ operators of bosons (electron-hole pairs) [41]

$$c^\dagger_\mathbf{p}=\frac{1}{\sqrt{S}}\int d\mathbf{r}_1 d\mathbf{r}_2\exp\left(\frac{i}{\hbar}\mathbf{p}\mathbf{R}_{12}\right)\psi^{(e)\dagger}_\uparrow(\mathbf{r}_1)\phi(\mathbf{r}_1,\mathbf{r}_2)\psi^{(h)\dagger}_\downarrow(\mathbf{r}_2), \tag{41}$$

$$c_\mathbf{p}=\frac{1}{\sqrt{S}}\int d\mathbf{r}_1 d\mathbf{r}_2\exp\left(-\frac{i}{2\hbar}\mathbf{p}\mathbf{R}_{12}\right)\psi^{(e)}_\uparrow(\mathbf{r}_1)\phi^*(\mathbf{r}_1,\mathbf{r}_2)\psi^{(h)}_\downarrow(\mathbf{r}_2), \tag{42}$$

where $S$ is the area of the system. The deviation of the commutation relations for the operators $c^\dagger_\mathbf{p}$ and $c_\mathbf{p}$ from the Bose commutation relations is of order $na_B^2$, so that in the low-density limit electron-hole pairs can be treated as bosons.

The term $H_T$, responsible for the tunneling of carriers through the barrier, now takes the form

$$H_T=\sum_{\mathbf{p},\mathbf{q}}\left(T_{\mathbf{p},\mathbf{q}}c^\dagger_{\mathbf{p},l}c_{\mathbf{q},r}+T^*_{\mathbf{p},\mathbf{q}}c^\dagger_{\mathbf{q},r}c_{\mathbf{p},l}\right)\equiv\Lambda+\Lambda^\dagger, \tag{43}$$

where $c^\dagger_{\mathbf{p},l}$ and $c_{\mathbf{q},r}$ are the creation and annihilation operators of an electron-hole pair to the left and right of the barrier, respectively, and $T_{\mathbf{p},\mathbf{q}}$ is the matrix element determining the tunneling probability of the pair. The tunneling current $I(t)$, for example, in the electron layer of the left region, is determined by the rate of change of the number of pairs in this region and is equal to

$$I(t)=e\left\langle\frac{d}{dt}\sum_\mathbf{p}c^\dagger_{\mathbf{p},l}c_{\mathbf{p},l}\right\rangle=\frac{2e}{\hbar}\operatorname{Im}e^{-i\varphi(t)}\left\langle\Lambda(t)\right\rangle. \tag{44}$$

Here the phase difference $\varphi(t)$ between the left and right regions is written explicitly. In the presence of an external total voltage $V(t)=V_e(t)+V_h(t)$, it is determined by the equation

$$\hbar\frac{d\varphi(t)}{dt}=eV(t). \tag{45}$$

Assuming weak tunneling, we use perturbation theory. In this case, the expression for the current (44), up to terms quadratic in $H_T$, takes the form

$$I(t) = I_0(t) + I_1(t) = \frac{2e}{\hbar}\operatorname{Im} e^{-i\varphi(t)} \left\langle \Lambda(t) \right\rangle_0 - \\ -\frac{2e}{\hbar^2}\operatorname{Re}\int_{-\infty}^{t} dt' \left\{ e^{-i(\varphi(t)+\varphi(t'))} \left\langle \left[ \Lambda(t), \Lambda(t') \right] \right\rangle_0 + e^{-i(\varphi(t)-\varphi(t'))} \left\langle \left[ \Lambda(t), \Lambda^\dagger(t') \right] \right\rangle_0 \right\}. \tag{46}$$

Here, as before, the symbol $\langle ... \rangle_0$ denotes the average taken with respect to the state of the unperturbed Hamiltonian.

A distinctive feature of the current arising in a system with a low carrier density is that this current does not vanish in the first order in $H_T$. Indeed, taking into account that most pairs are in the condensate and, accordingly, neglecting transitions from noncondensed pairs to noncondensed ones, the operator $\Lambda$ can be decomposed into three terms [42]

$$\Lambda = T_{00} c_{0,l}^\dagger c_{0,r} + \sum_{\mathbf{p}\neq 0} T_{\mathbf{p}0} c_{\mathbf{p},l}^\dagger c_{0,r} + \sum_{\mathbf{q}\neq 0} T_{0\mathbf{q}} c_{0,l}^\dagger c_{\mathbf{q},r}, \tag{47}$$

where the first term describes transitions of condensed pairs from the right region to condensed pairs in the left region, the second term corresponds to transitions of condensed pairs from the right region to noncondensed pairs in the left region, and the third term corresponds to transitions of noncondensed pairs from the right region to condensed pairs in the left region. Since $c_{0,l}$ and $c_{0,r}$ are the average values of the operators $c_{\mathbf{p},l}$ and $c_{\mathbf{q},r}$, and are equal to $\sqrt{N_l}$ and $\sqrt{N_r}$, where $N_l$ and $N_r$ are the average numbers of pairs to the left and right of the barrier, respectively, we obtain from (46)

$$I_0(t) = -\frac{2eT_{00}}{\hbar}\sqrt{N_l N_r}\sin\varphi(t) \equiv I_c \sin\varphi(t). \tag{48}$$

In the second order in $H_T$ the current acquires an additional contribution $I_1(t)$, whose calculation is analogous to the calculation of the current (8), corresponding to the case of high carrier density. Similarly to (21), the current $I_1(t)$ consists of a quasiparticle (normal) current $I_q$, as well as superconducting and interference currents with amplitudes $I_{J1}$ and $I_{J2}$, and is given by

$$I_1(t) = I_q\left(V(t)\right) + I_{J1}\left(V(t)\right)\sin 2\varphi(t) + I_{J2}\left(V(t)\right)\cos 2\varphi(t). \tag{49}$$

The amplitudes $I_q$, $I_{J1}$ and $I_{J2}$ are determined by the following expressions:

$$I_q(V) = -\frac{eN_l}{\hbar^2}\sum_{\mathbf{q}} \left| T_{0\mathbf{q}} \right|^2 A_b\left(q, -\frac{eV}{\hbar}\right) + \frac{eN_r}{\hbar^2}\sum_{\mathbf{p}} \left| T_{\mathbf{p}0} \right|^2 A_b\left(p, \frac{eV}{\hbar}\right), \tag{50}$$

$$I_{J2}(V) = -\frac{eN_l}{\hbar^2}\sum_{\mathbf{q}} \left| T_{0\mathbf{q}} \right|^2 B_b\left(q, \frac{eV}{\hbar}\right) - \frac{eN_r}{\hbar^2}\sum_{\mathbf{p}} \left| T_{\mathbf{p}0} \right|^2 B_b\left(p, \frac{eV}{\hbar}\right), \tag{51}$$

$$I_{J1}(V) = -P\int_{-\infty}^{+\infty} \frac{d\omega}{\pi} \frac{I_{J2}(\hbar\omega / e)}{\omega + eV/\hbar}. \tag{52}$$

Here $A_b(k,\omega)$ and $B_b(k,\omega)$ are the normal and anomalous spectral functions of the weakly interacting homogeneous Bose gas

$$A_b(k,\omega) = \pi\left( \left( \frac{\varepsilon_k + \mu}{E_k} + 1 \right)\delta(\omega - E_k/\hbar) - \left( \frac{\varepsilon_k + \mu}{E_k} - 1 \right)\delta(\omega + E_k/\hbar) \right), \tag{53}$$

$$B_b(k,\omega) = \frac{\pi\mu}{E_k}\left( \delta(\omega + E_k/\hbar) - \delta(\omega - E_k/\hbar) \right), \tag{54}$$

where $\varepsilon_k = k^2/2m$, and $E_k$ is the spectrum of the collective mode $E_k = \sqrt{\varepsilon_k(\varepsilon_k + 2\mu)}$. Explicit expressions for $I_c$, $I_q$, $I_{J1}$ and $I_{J2}$ can be obtained after substituting into (48), (50)-(52) the matrix elements $T_{00}$, $T_{\mathbf{p}0}$ and $T_{0\mathbf{q}}$, which are expressed through the wave functions $\Psi$, obtained from the solutions of the Gross–Pitaevskii equation (38) and the Bogoliubov–de Gennes equation.

The key difference between the amplitudes (50)-(52), obtained for the case of low carrier density, and expressions (25)-(27), corresponding to the high-density regime, is that the former do not depend on temperature. In particular, this means that even at zero temperature the amplitudes $I_q$ and $I_{J2}$ remain nonzero, leading to damping of the phase $\varphi(t)$ and, consequently, of the Josephson oscillations in the case of a constant voltage $V(t) = V_0$ [42]. Indeed, taking into account that the phase difference $\varphi(t)$ between the right and left sides is determined by the difference of chemical potentials $\mu_l(\bar{N}+\nu) - \mu_r(\bar{N}-\nu) = \Delta\mu$, where $\bar{N} = (N_l + N_r)/2$ is the average number of pairs, and $\nu = (N_l - N_r)/2$ – is the pair imbalance, and assuming that $\nu << \bar{N}$, to linear order in $\nu$ instead of (45) we obtain

$$\hbar \frac{d\varphi(t)}{dt} = \Delta\mu = eV_0 - U\nu \ . \tag{55}$$

Here $U = \partial\mu_l / \partial N_l + \partial\mu_r / \partial N_r$ is an energy of the order of $\bar{\mu}/\bar{N}$. Thus, the phase difference is determined both by the external voltage $eV_0 = \mu_l(\bar{N}) - \mu_r(\bar{N})$, and by the pair imbalance $\nu$. Differentiating expression (55) with respect to time and taking into account that $e\dot{\nu}$ is the total current $I(t)$, we obtain

$$\frac{e\hbar}{U}\frac{d^2\varphi(t)}{dt^2} + I_c \sin\varphi(t) + I_q(\Delta\mu) + I_{J1}(\Delta\mu)\sin 2\varphi(t) + I_{J2}(\Delta\mu)\cos 2\varphi(t) = 0 \ . \tag{56}$$

We further restrict ourselves to the case of small $\Delta\mu$. As follows from (51), (52) and (54), in the limit $\Delta\mu \to 0$ the amplitude $I_{J1}$ remains finite and is equal to

$$I_{J1}(0) = -\frac{2}{\hbar}\left( N_l\mu_r \sum_{\mathbf{q}} \frac{|T_{0\mathbf{q}}|^2}{E_q^2} + N_r\mu_l \sum_{\mathbf{p}} \frac{|T_{\mathbf{p}0}|^2}{E_p^2} \right). \tag{57}$$

The amplitudes $I_q$ and $I_{J2}$, on the other hand, decrease linearly with $\Delta\mu$, i.e., $I_q(\Delta\mu \to 0) = I_{J2}(\Delta\mu \to 0) = \Delta\mu / eR = \hbar\dot{\varphi} / eR$, where $R$ is the resistance associated with the quasiparticle current. This result is related to the fact that for small $\Delta\mu$ the main contribution to the expressions (50) and (51) is provided by low-energy phonon excitations (see (53) и (54)), leading to a gapless transport. As a result, Eq. (56) takes the form

$$\frac{e\hbar}{U}\frac{d^2\varphi(t)}{dt^2} + \frac{\hbar}{eR}\left(1 + \cos 2\varphi(t)\right)\frac{d\varphi(t)}{dt} + I_c \sin\varphi(t) + I_{J1}(0)\sin 2\varphi(t) = 0 \ , \tag{58}$$

which is analogous to the RCSJ model [37], where the first term plays the role of the displacement current $I_D = (\hbar C / e)\ddot{\varphi}(t)$ through the capacitance $C = e^2 / U$ (although this current does not actually pass through the weak contact, for the external circuit containing the Josephson junction it effectively adds to the other current components). From this it can be seen that the presence of currents $I_q$ and $I_{J2}$ in the system under consideration leads to damping of the Josephson oscillations even at zero temperature. This result is fundamentally different from the result in a system with high carrier density, where at zero temperature the presence of an energy gap in the quasiparticle spectrum leads to a step-like dependence of the currents $I_q$ and $I_{J2}$ on the potential difference: when $\Delta\mu$ is below a certain threshold value, these currents vanish and only the supercurrent flows in the system.

The degree of damping of the Josephson oscillations is not difficult to obtain at times close to the moment when the voltage $V_0$ is switched on. In this case, the phase remains small $\varphi(t \to 0) = eV_0 t / \hbar$ (the initial phase difference is assumed to be zero) and the oscillatory terms in (58) can be linearized. As a result, the equation (58) reduces to a linear damped oscillator equation for $\varphi(t)$. As is well known, its solution describes damped oscillations whose amplitude decays exponentially in time as $\exp(-t/\tau)$, where $\tau = RC$. The regime of weak damping, in which the amplitude changes slightly over one oscillation period, corresponds to the condition $1/\tau^2 << eI_c/\hbar C$.

If a small oscillating component $\delta V = V_a \sin\omega t$ is added to the applied constant voltage $V_0$ an additional term $-e^2 V_a \omega \cos\omega t / U$ appears on the right-hand side of Eq. (58). This term plays the role of an external force determining the dynamics of the phase $\varphi$. Under resonance conditions $\omega = neV_0/\hbar$ ($n = 1,2,3...$), a time-averaged supercurrent will arise in the system through the weak link (the Shapiro effect).

As noted in [21], in the case where the distance between the barriers in the longitudinal direction is much larger than the coherence length (see Fig. 2), the phase of the order parameter undergoes a jump at each of the barriers.

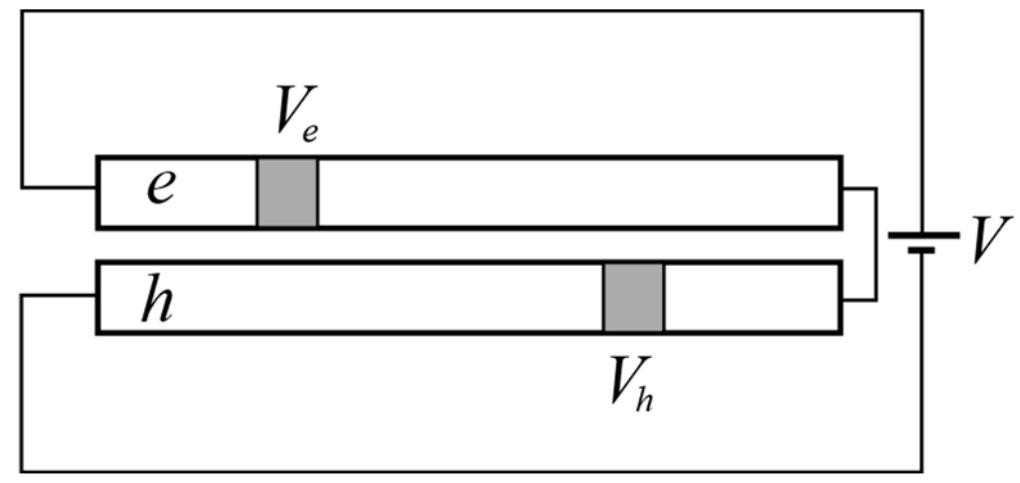

*Fig. 2. Schematic representation of a bilayer system in which the barriers in the electron and hole layers are separated from each other in the longitudinal direction.*

In the presence of a constant external voltage $V_0$, the supercurrent through each of the barriers is determined by (48):

$$I(t) = I_{c1} \sin\varphi_1(t) = I_{c2} \sin\varphi_2(t), \tag{59}$$

where $I_{c1,2}$ and $\varphi_{1,2}(t)$ are the critical currents and the phase jumps at each of the barriers, respectively. The total phase difference $\varphi(t) = \varphi_1(t) + \varphi_2(t)$ satisfies Eq. (45) and is given by $\varphi(t) = eV_0 t/\hbar + \varphi_0$ ($\varphi_0$ is a constant). Substituting $\varphi_2(t) = \varphi(t) - \varphi_1(t)$ into Eq. (59) and solving this equation for $\varphi_1$, we obtain expressions for the phase jumps at each of the barriers

$$\varphi_1(t) = \arctan\left(\frac{I_{c2}\sin\varphi(t)}{I_{c1} + I_{c2}\cos\varphi(t)}\right), \ \varphi_2(t) = \varphi(t) - \varphi_1(t), \tag{60}$$

as well as the expression for the supercurrent

$$I(t) = \frac{I_{c1} I_{c2} \sin\varphi(t)}{\sqrt{I_{c1}^2 + I_{c2}^2 + 2I_{c1}I_{c2}\cos\varphi(t)}}. \tag{61}$$

Since the voltage drops across the barriers in the electron layer $V_e$ and the hole layer $V_h$ are related to the time derivatives of the phases $\varphi_1(t)$ and $\varphi_2(t)$ through relation (45), these voltages oscillate according to the law

$$V_e(t) = V_0 \frac{I_{c2}\left(I_{c2} + I_{c1}\cos\varphi(t)\right)}{I_{c1}^2 + I_{c2}^2 + 2I_{c1}I_{c2}\cos\varphi(t)}, \ V_h(t) = V_0 - V_e(t). \tag{62}$$

The voltage drops become constants ($V_e = V_h = V_0/2$) in the case of identical barriers (and, correspondingly, equal critical currents $I_{c1,2}$).

It should be noted that the results presented above were obtained neglecting the resistive quasiparticle and interference currents, whose magnitudes arise in second order in $H_T$, whereas the supercurrent is obtained already in first order in $H_T$. In addition, capacitive effects associated with the relative particle number $\nu$ were also neglected. Taking them into account leads to a slight deviation of the supercurrent oscillations from a purely sinusoidal form [42], but does not qualitatively affect the result.

## Appendix

In this Appendix, we demonstrate the connection between the Keldysh wave function (36), coherent states, and the Bardeen–Cooper–Schrieffer (BCS) wave function.

As is well known, the coherent state of a dilute Bose gas $\left|\Phi_B\right\rangle$ is an eigenstate of the field annihilation operator $\hat{\Psi}(\mathbf{R})$, i.e.,

$$\hat{\Psi}(\mathbf{R})\left|\Phi_B\right\rangle = \Psi(\mathbf{R})\left|\Phi_B\right\rangle, \tag{A-1}$$

where $\Psi(\mathbf{R})$ is the complex order parameter. The explicit form of this state is known (see, e.g., [12]):

$$\left|\Phi_B\right\rangle = \exp\left(\int \Psi(\mathbf{R})\hat{\Psi}^\dagger(\mathbf{R})d\mathbf{R} - h.c.\right)\left|0\right\rangle. \tag{A-2}$$

At low densities, electron-hole pairs behave, to zeroth order, as a weakly interacting Bose gas. Therefore, the coherent state (A-2) can also be applied to a dilute exciton gas. To account for internal degrees of freedom, we write the expansion

$$\hat{\Psi}^\dagger(\mathbf{R}) = \frac{1}{\sqrt{S}}\sum_{\mathbf{p}} \hat{c}^\dagger_{\mathbf{p}} \exp\left(-i\mathbf{p}\mathbf{R}/\hbar\right), \tag{A-3}$$

where $\hat{c}^\dagger_{\mathbf{p}}$ is defined by (41). Substituting (41) into (A-3), and then into (A-2), we obtain

$$\left|\Phi_B\right\rangle = \exp\left(\int d\mathbf{r}_1 d\mathbf{r}_2 \psi^{(e)\dagger}_{\uparrow}(\mathbf{r}_1)\Psi(\mathbf{R}_{12})\phi\left(\mathbf{r}_1,\mathbf{r}_2\right)\psi^{(h)\dagger}_{\downarrow}(\mathbf{r}_2) - h.c.\right)\left|0\right\rangle. \tag{A-4}$$

Returning to the Keldysh wave function (36), written for the case of singlet pairing and phase separation in the form

$$\left|\Phi_K\right\rangle = \exp\left\{\int d\mathbf{r}_1 d\mathbf{r}_2 \Phi(\mathbf{r}_1,\mathbf{r}_2)\psi^{(e)\dagger}_{\uparrow}(\mathbf{r}_1)\psi^{(h)\dagger}_{\downarrow}(\mathbf{r}_2) - h.c.\right\}\left|0\right\rangle, \tag{A-5}$$

we note, that due to the completeness of the set of eigenfunctions $\phi_n(\mathbf{r})$ of the Hamiltonian $H_{\text{int}}(\mathbf{r})$ describing the internal degrees of freedom, the function $\Phi(\mathbf{r}_1,\mathbf{r}_2)$ can be represented as the expansion

$$\Phi(\mathbf{r}_1,\mathbf{r}_2) = \sum_n \phi_n(\mathbf{r}_{12})\Psi_n(\mathbf{R}_{12}). \tag{A-6}$$

Since the energy spacing between neighboring states $\phi_n$ and $\phi_{n+1}$ is finite, at zero temperature it is sufficient to retain only the first term in the expansion (A-6). As a result, expression (A-5) reduces to (A-4). Thus, in the low-density limit $na_B^2 << 1$, where deviations of exciton statistics from Bose statistics are small and the expression for the operator $\hat{c}^\dagger_{\mathbf{p}}$ given by (41) is valid, the Keldysh wave function reduces to the coherent state of bosons.

It is also useful to note that in the spatially homogeneous case ($\Phi(\mathbf{r}_1,\mathbf{r}_2) = \Phi(\mathbf{r}_{12})$), the wave function (A-5) reduces to a BCS-like wave function [43]. Indeed, after performing the Fourier transforms

$$\Phi(\mathbf{r}_{12}) = \frac{1}{S}\sum_{\mathbf{k}} \Phi_{\mathbf{k}} e^{i\mathbf{k}(\mathbf{r}_1-\mathbf{r}_2)},\ \psi^{(e)\dagger}_{\uparrow}(\mathbf{r}_1) = \frac{1}{\sqrt{S}}\sum_{\mathbf{q}} a^\dagger_{\mathbf{q}} e^{i\mathbf{q}\mathbf{r}_1},\ \psi^{(h)\dagger}_{\downarrow}(\mathbf{r}_2) = \frac{1}{\sqrt{S}}\sum_{\mathbf{p}} b^\dagger_{\mathbf{p}} e^{i\mathbf{p}\mathbf{r}_2}, \tag{A-7}$$

and substituting (A-7) into (A-5), we obtain

$$\left|\Phi_K\right\rangle = \exp\left(\sum_{\mathbf{k}} \Phi_{\mathbf{k}} a^\dagger_{\mathbf{k}} b^\dagger_{-\mathbf{k}} - h.c.\right)\left|0\right\rangle = \prod_{\mathbf{k}} \exp\left(\Phi_{\mathbf{k}} a^\dagger_{\mathbf{k}} b^\dagger_{-\mathbf{k}} - h.c.\right)\left|0\right\rangle \equiv \prod_{\mathbf{k}} \exp\left(\hat{D}_{\mathbf{k}}\right)\left|0\right\rangle. \tag{A-8}$$

Taking into account that $\hat{D}_{\mathbf{k}}^{2n+1}=(-1)^n\left|\Phi_{\mathbf{k}}\right|^{2n}\hat{D}_{\mathbf{k}}$, where $n=0,1,2...$, and expanding the exponential function in (A-8) into a Taylor series, we finally arrive at

$$\left|\Phi_K\right\rangle=\prod_{\mathbf{k}}\left(u_{\mathbf{k}}+v_{\mathbf{k}}a_{\mathbf{k}}^{\dagger}b_{-\mathbf{k}}^{\dagger}\right)\left|0\right\rangle, \quad \text{(A-9)}$$

where $u_{\mathbf{k}}=\cos|\Phi_{\mathbf{k}}|$, $v_{\mathbf{k}}=(\Phi_{\mathbf{k}}/|\Phi_{\mathbf{k}}|)\sin|\Phi_{\mathbf{k}}|$. Thus, in the spatially homogeneous case, the Keldysh wave function coincides with the BCS wave function.